\documentclass[pra,twocolumn,showpacs,floatfix,amsmath,amsfonts]{revtex4}
\usepackage{graphicx}

\newcommand{\bra}[1]{\langle #1 | \,}
\newcommand{\ket}[1]{\, | #1 \rangle}

\newcommand{\expv}[1]{\langle #1 \rangle}
\newcommand{\ddt}{\frac{\partial}{\partial t}}
\newcommand{\ddz}{\frac{\partial}{\partial z}}
\newcommand{\lra}{\leftrightarrow}

\newcommand{\La}{\Lambda}
\newcommand{\om}{\omega}
\newcommand{\Om}{\Omega}
\newcommand{\ga}{\gamma}
\newcommand{\Ga}{\Gamma}
\newcommand{\de}{\delta}
\newcommand{\De}{\Delta}
\newcommand{\ka}{\kappa}
\newcommand{\sih}{\hat{\sigma}}
\newcommand{\phih}{\hat{\phi}}
\newcommand{\Eh}{\hat{\cal E}}
\newcommand{\Fh}{\hat{F}}
\newcommand{\Ih}{\hat{I}}
\newcommand{\Fch}{\hat{\cal F}}
\newcommand{\zp}{z^{\prime}}
\newcommand{\zpp}{z^{\prime \prime}}
\newcommand{\eps}{\epsilon}
\newcommand{\veps}{\varepsilon}

\begin{document}

\title{Magneto-optical rotation and cross-phase modulation 
via coherently driven tripod atoms}

\author{David Petrosyan}
\affiliation{Institute of Electronic Structure \& Laser, 
FORTH, Heraklion 71110, Crete, Greece}
\author{Yuri P. Malakyan}
\affiliation{Institute for Physical Research, NAS of Armenia,
Ashtarak-2, 378410, Armenia}

\date{\today}

\begin{abstract}
We study the interaction of a weak probe field, having two orthogonally 
polarized components, with an optically dense medium of four-level atoms 
in a tripod configuration. In the presence of a coherent driving laser, 
electromagnetically induced transparency is attained in the medium,
dramatically enhancing its linear as well as nonlinear dispersion 
while simultaneously suppressing the probe field absorption. 
We present the semiclassical and fully quantum analysis of the system. We 
propose an experimentally feasible setup that can induce large Faraday 
rotation of the probe field polarization and therefore be used for 
ultra-sensitive optical magnetometry. We then study the Kerr nonlinear 
coupling between the two components of the probe, demonstrating a novel 
regime of symmetric, extremely efficient cross-phase modulation, capable 
of fully entangling two single-photon pulses. This scheme may thus pave 
the way to photon-based quantum information applications, such as 
deterministic all-optical quantum computation, dense coding and teleportation. 
\end{abstract}


\maketitle

\section{Introduction}

Electromagnetically induced transparency (EIT) in atomic media is 
a quantum interference effect that results in a dramatic reduction 
of the group velocity of propagating probe field accompanied by 
vanishing absorption \cite{eit_rev,ScZub,vred}. As the quantum 
interference is usually very sensitive to the system parameters, 
various schemes exhibiting EIT are attracting growing attention 
in view of their potential for significantly enhancing nonlinear 
optical effects. Some of the most representative examples include 
slow-light enhancement of acusto-optical interactions in doped fibers 
\cite{acopt}, trapping light in optically dense atomic and doped
solid-state media by coherently converting photonic excitation into 
spin excitation \cite{fllk,v0exp,hemmer} or by creating a photonic band gap 
via periodic modulation of the EIT resonance \cite{lukin-pbg}, 
and nonlinear photon-photon coupling using N configuration of 
atomic levels \cite{imam,harris}.

EIT is based on the phenomenon of coherent population trapping 
\cite{eit_rev,ScZub}, in which the application of two laser fields 
to a three-level $\La$ system creates the so-called ``dark state'', 
which is stable against absorption of both fields. Dark states are 
also found in several other multilevel systems, one of them being 
four-level atoms interacting with three laser fields in tripod 
configuration. Tripod atoms proved to be robust systems for ``engineering''
arbitrary coherent superpositions of atomic states \cite{bergmann} 
using an extension of the well-known technique of stimulated Raman 
adiabatic passage (STIRAP) \cite{bergmann-rev}. Parametric generation 
of light in a medium of tripod atoms, prepared in a certain coherent 
superposition of ground states, has been recently discussed in 
\cite{pasp-knight}. In a related work, it was shown that enhanced
nonlinear conversion between two laser pulses is attainable in
a medium of $\La$ atoms with spatially dependent ground state 
coherence \cite{kis-pasp}. In the present paper we undertake a detailed 
study of propagation of a weak probe field through a medium of tripod 
atoms under the conditions of EIT \cite{yumal}. We show that this system 
can support large magneto-optical rotation (MOR) of the probe field 
polarization, accompanied by negligible absorption. It can 
therefore be used for ultra-sensitive optical magnetometry, 
with the sensitivity comparable to (or better than) other hitherto 
studied MOR schemes \cite{budker_rev}. In contrast to these schemes, 
where the basic mechanism of nonlinear MOR is the probe field induced 
coherence between the Zeeman sublevels of atomic ground state 
\cite{butker,scully-mor}, in our case the MOR results from an 
extraordinary dispersion induced by a strong driving field in the 
EIT regime. Hence, by simply changing the intensity of the driving 
field, one could control the polarization rotation of the 
weak probe field. We note that an interferometric measurement of the 
magnetic field induced phase shift of the probe, subject to EIT in 
the presence of a driving field, can yield sensitivity of the order 
of $10^{-12}$ G \cite{flscl}. These studies and our present 
contribution reveal the significant potential for improving the 
sensitivity of Faraday magnetometers to small magnetic fields as 
compared to conventional optical pumping magnetometers \cite{opt-pump}.

Another motivation for the present work is its relevance to the 
field of quantum information (QI), which is attracting broad interest 
in view of its fundamental nature and its potentially revolutionary 
applications to cryptography, teleportation and computing \cite{QCQI}. 
Among the various QI processing schemes of current interest 
\cite{solst,iontr,BCJD,linopt,phphcav}, those based on photons 
\cite{linopt,phphcav} have the advantage of using very robust and 
versatile carriers of QI. Yet the main impediment towards their 
operation at the few-photon level is the weakness of optical nonlinearities 
in conventional media \cite{Boyd}. As mentioned above, EIT schemes with 
atoms having N configuration of levels have opened up a possibility
of achieving enhanced nonlinear coupling of weak quantum fields at the
single-photon level \cite{imam,harris}. The main hindrance of such 
schemes is the mismatch between the group velocities of the pulse 
subject to EIT and its nearly-free propagating partner, which severely 
limits their effective interaction length \cite{harris}. This drawback 
may be remedied by using an equal mixture of two isotopic species, 
interacting with two driving fields and an appropriate magnetic field, 
which would render the group velocities of the two pulses equal \cite{lukimam}.
Here we propose an alternative, simple and robust approach which relies 
solely on an intra-atomic process, without resorting to two isotopic species 
and using just one driving field \cite{yumal,rebic}. In our scheme, two 
orthogonally polarized weak (quantum) fields, acting on adjacent transitions 
of tripod atoms, propagate with the same group velocity and impress large 
conditional phase shift upon each other. 

The paper is organized as follows. In Sec. \ref{form} we formulate the 
theory and give an analytical solution of the equations of motion for the 
two components of the weak probe field. In Sec. \ref{mor} we discuss the 
setup and sensitivity limits of the optical magnetometer. Section \ref{xpm} 
is devoted to the study of feasibility of strong nonlinear interaction 
and entanglement between two orthogonally polarized weak quantum fields, 
aimed at quantum information applications. Our conclusions are summarized 
in Sec. \ref{sum}.

\section{Formulation}
\label{form}

\begin{figure}[t]
\centerline{\includegraphics[width=8.5cm]{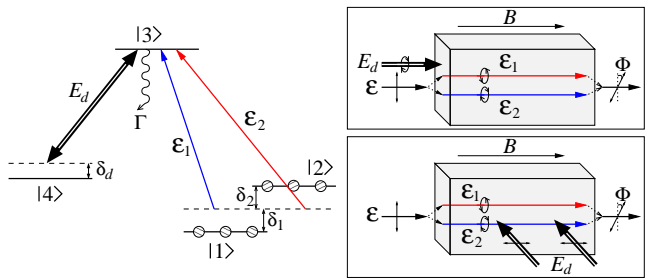}}
\caption{Level scheme of tripod atoms interacting with weak 
probe ${\cal E}$ and strong driving $E_d$ fields.  
Upper inset: copropagating probe with circularly left- and right-polarized 
components ${\cal E}_{1,2}$, and driving $E_d$ fields pass through the 
atomic medium that is subject to the longitudinal magnetic field $B$. 
Lower inset: Perpendicular arrangement of the probe and driving fields,
that is suitable for cold atomic gas.}
\label{as}
\end{figure}

We consider a near-resonant interaction of two optical fields with a 
medium of atoms with tripod configuration of levels (Fig. \ref{as}). 
The medium is subject to a longitudinal magnetic field $B$ that 
removes the degeneracy of the ground state sublevels. The Zeeman 
shift of levels $\ket{1}$ and $\ket{2}$ is given by 
$\hbar \De = \mu_{\rm B} M_F g_F B$, where $\mu_{\rm B}$ is the Bohr 
magneton, $g_F$ is the gyromagnetic factor and $M_F = \pm 1$ is the 
magnetic quantum number of the corresponding state.  All the atoms 
are assumed to be optically pumped to the states $\ket{1}$ and $\ket{2}$ 
which thus have the same incoherent populations equal to $1/2$.
A linearly polarized weak (quantum) probe field ${\cal E}$ has a carrier 
frequency $\om_p$ and wavevector $k_p$ parallel to the magnetic field 
direction. Its two circularly left- and right-polarized components 
${\cal E}_{1,2}$ act on the atomic transitions $\ket{1} \to \ket{3}$ and  
$\ket{2} \to \ket{3}$, with the detunings 
$\de_{1,2} = \om_p - \om_{31}^{0} - k_p v \mp \De$, where 
$\om_{31}^{0} = \om_{32}^{0}$ is the frequency of the unshifted 
atomic resonance and $k_p v$ is the Doppler shift for the atoms having 
velocity $v$ along the probe field propagation direction. A strong 
classical cw field $E_d$, having frequency $\om_d$ and wavevector 
$k_d \simeq k_p$, is driving the atomic transition 
$\ket{3} \lra \ket{4}$ with the Rabi frequency $\Om_d= \wp_{34} E_d/\hbar$, 
where $\wp_{\mu \nu}$ is the dipole matrix element on the transition 
$\ket{\mu} \to \ket{\nu}$. In the collinear Doppler-free geometry shown 
in Fig. \ref{as}, upper inset, the driving field has to be circularly left 
or right polarized, in order to couple to a single magnetic sublevel 
$\ket{4}$. Its Zeeman shift 
$\hbar \De^{\prime} = \mu_{\rm B} M_{F^{\prime}} g_{F^{\prime}} B$ 
is incorporated in the detuning of the driving field via 
$\de_d=\om_d-\om_{34}^{0}- k_d v +\De^{\prime}$, where $\om_{34}^{0}$ 
is the atomic resonance frequency for zero magnetic field. Note that in 
the case of cold atomic sample (Doppler broadening of the atomic resonance 
is smaller than the ground-state spin relaxation rate), one can employ
the perpendicular geometry of  Fig. \ref{as}, lower inset, where the
driving field is linearly $\pi$ polarized while the Zeeman shift of
level $\ket{4}$ vanishes, $\De^{\prime}=0$ since $M_{F^{\prime}}=0$.

\begin{figure}[t]
\centerline{\includegraphics[width=8.5cm]{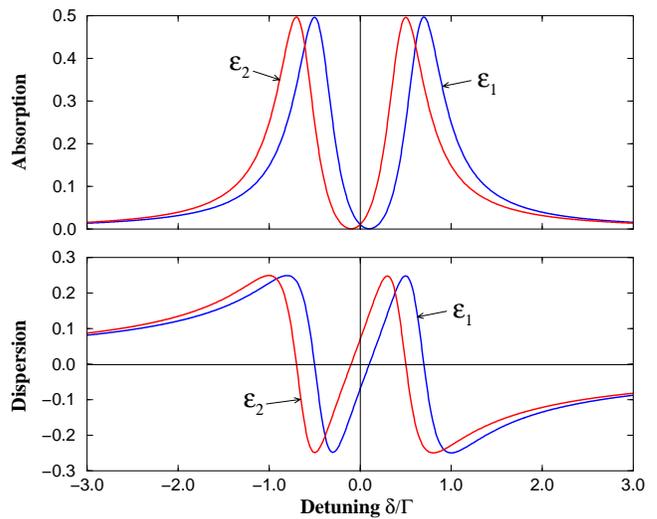}}
\caption{Absorption and dispersion spectra ($\de = \om_p - \om_{31}^0$) 
for the ${\cal E}_1$ and ${\cal E}_2$ components of the probe in the 
presence of a strong driving ($\Om_d =0.6\Ga$) and a weak magnetic 
($\De = 0.1 \Ga$) fields, in units of the linear resonant absorption 
coefficient $a_0$.}
\label{abdis}
\end{figure}

To illustrate the scheme, we plot in Fig. \ref{abdis} the absorption and 
dispersion spectra of the two components of the probe field ${\cal E}$ for 
the case $\de_d=0$. In the presence of magnetic field, the spectra for 
the ${\cal E}_1$ and ${\cal E}_2$ are {\it shifted} with respect to each 
other by the amount equal to the Zeeman shift $2\De$. When the probe field 
is resonant with the unshifted ($\De=0$) atomic transitions, 
$\om_p = \om_{31}^{0}= \om_{32}^{0}$, due to the steep and 
approximately linear slope of the dispersion in the vicinity of 
$\de_{1,2} = 0$, upon propagating through the medium the two 
components of the probe experience equal and opposite phase shifts 
$\phi_1 = -\phi_2$ which results in a net polarization rotation 
of the field, $\Phi = \frac{1}{2}(\phi_2 -\phi_1)$. If the Zeeman 
shift is small compared to the width of the EIT window for both 
components of the probe, the absorption remains much smaller than 
the phase shift. Thus, a weak magnetic field can induce an appreciable 
polarization rotation accompanied by negligible absorption, allowing 
for extremely sensitive magnetometry (Sec. \ref{mor}). 
In addition to the large linear phase shift, each component 
experiences a nonlinear cross-phase modulation. Although this 
cross-phase modulation is typically small compared to the linear 
phase modulation, it is nevertheless several orders of magnitude 
larger than that in conventional media \cite{imam}. It can therefore 
be used for quantum information applications based on photon-photon 
interaction and entanglement (Sec. \ref{xpm}).

Let us now consider the scheme more quantitatively. We describe 
the medium using collective slowly varying atomic operators 
$\sih_{\mu \nu}(z,t)=\frac{1}{N_z} \sum_{j=1}^{N_z} \ket{\mu_j}\bra{\nu_j}$, 
averaged over small but macroscopic volume containing many atoms 
$N_z = (N/L) dz \gg 1$ around position $z$, where $N$ is the total 
number of atoms and $L$ is the length of the medium \cite{fllk}. 
The two components of the quantum probe field are described by 
the corresponding field operators $\Eh_{1,2}$. 
In a frame rotating with the probe and driving field frequencies, 
the interaction Hamiltonian has the following form
\begin{eqnarray}
H &=& \hbar \frac{N}{L}\int_0^L dz 
[\de_1 \sih_{11} + \de_2 \sih_{22} + \de_d \sih_{44}
\nonumber \\ & & \;\;\;\;\; \;\;\;\;\; 
-g (\Eh_1 \sih_{31} + \Eh_2 \sih_{32}) - \Om_d \sih_{34}
 + {\rm H. c.}].
\label{ham}
\end{eqnarray}
Here $g = \wp_{31} \sqrt{ \om_p /(2 \hbar \eps_0 AL)}$, with $A$ being the 
cross-sectional area of the probe field, is the atom-field coupling constant, 
which is the same for both circular components $\Eh_{1,2}$ due to the 
symmetry of the system ($|\wp_{31}| = |\wp_{32}|$ while the opposite signs 
of the Clebsch-Gordan coefficients on the transitions $\ket{1} \to \ket{3}$ 
and $\ket{2} \to \ket{3}$ can be incorporated into the atomic eigenstate 
via the transformation $\ket{1} \to e^{i \pi} \ket{1}$).
Using the slowly varying envelope approximation, we obtain the following 
propagation equations for the quantum field operators
\begin{subequations}
\label{Es}
\begin{eqnarray}
\left(\ddt + c \ddz \right) \Eh_1(z,t) &=& 
i g N \sih_{13},\label{E1} \\
\left(\ddt + c \ddz \right) \Eh_2(z,t) &=& 
i g N \sih_{23} \label{E2}. 
\end{eqnarray}
\end{subequations}
The equations for the atomic coherences are given by 
\begin{widetext}
\begin{subequations}
\label{ss}
\begin{eqnarray}
\ddt \sih_{12} &=& 
[i (\de_1 - \de_2) - \ga_c ]\sih_{12} - i g \Eh_1 \sih_{32} 
+ i g \Eh_2^{\dagger} \sih_{13} + \Fh_{12} , \label{s12}\\
\ddt \sih_{13} &=&
\left(i \de_1 - \frac{\Ga}{2}\right) \sih_{13} 
+ i g \Eh_1 (\sih_{11}- \sih_{33})
+ i g \Eh_2 \sih_{12} + i \Om_d \sih_{14} + \Fh_{13} , \label{s13}\\
\ddt \sih_{14} &=&
[i (\de_1 - \de_d) - \ga_c ]\sih_{14} - i g \Eh_1 \sih_{34}
+ i \Om_d^{*} \sih_{13} + \Fh_{14} , \label{s14}\\
\ddt \sih_{23} &=&
\left(i \de_2 - \frac{\Ga}{2}\right) \sih_{23} 
+ i g \Eh_2 (\sih_{22}- \sih_{33})
+ i g \Eh_1 \sih_{21} + i \Om_d \sih_{24} + \Fh_{23} , \label{s23}\\
\ddt \sih_{24} &=&
[i (\de_2 - \de_d) - \ga_c ]\sih_{24} - i g \Eh_2 \sih_{34}
+ i \Om_d^{*} \sih_{23} + \Fh_{24} ,\label{s24} \\
\ddt \sih_{34} &=&
- \left(i \de_d + \frac{\Ga}{2}\right) \sih_{34}
- i g \Eh_1^{\dagger} \sih_{14}
- i g \Eh_2^{\dagger} \sih_{24}
+ i \Om_d^{*} (\sih_{33} - \sih_{44})  + \Fh_{34} ,\label{s34}
\end{eqnarray}
\end{subequations}
\end{widetext}
where $\ga_c$ is the ground state coherence (spin) relaxation rate,
$\Ga$ is the decay rate of the excited state $\ket{3}$ and $\Fh_{\mu \nu}$
are $\de$-correlated noise operators associated with the relaxation.

We now outline the solution of Eqs. (\ref{ss}) in the weak field limit. 
To this end, we assume that the Rabi frequencies $g {\cal E}_{1,2}$ of 
the quantum fields are much smaller than $\Om_d$ and the number of photons 
in $\Eh_{1,2}$ is much less than the number of atoms, therefore 
$\sih_{11} = \sih_{22} \simeq {\mathbf 1}/2$ while 
$\sih_{33} = \sih_{44} = \sih_{34} \simeq 0$. We may thus treat the 
atomic equations perturbatively in the small parameters $g \Eh_{1,2}/\Om_d$.
In the first order, from (\ref{s13}) and (\ref{s23}) we have
\[
\sih_{14}^{(1)} = - \frac{g \Eh_1 }{2 \Om_d}, 
\;\;\;
\sih_{24}^{(1)} = - \frac{g \Eh_2 }{2 \Om_d}.
\]
Substituting these into (\ref{s14}) and (\ref{s24}), and neglecting for 
now the spin relaxation, we obtain
\begin{eqnarray*}
\sih_{13}^{(1)} &=& \left[ \ddt - i(\de_1-\de_d) \right] 
\frac{i g \Eh_1 }{2 |\Om_d|^2} 
\simeq \frac{g \Eh_1 (\de_1 - \de_d)}{2 |\Om_d|^2}, \\
\sih_{23}^{(1)} &=& \left[ \ddt - i(\de_2-\de_d) \right] 
\frac{i g \Eh_2 }{2 |\Om_d|^2} 
\simeq \frac{g \Eh_2 (\de_2 - \de_d)}{2 |\Om_d|^2} .
\end{eqnarray*}
In these equations, the last equalities result from the adiabatic 
approximation, i.e., we assume that the probe pulse changes slowly
enough so that the atoms follow the field adiabatically. 
Quantitatively, the adiabatic evolution requires that the rate
of change of the probe field, 
$\max[\partial_t {\cal E}/{\cal E}] \sim T_p^{-1}$, where $T_p$ is 
the temporal width of the pulse, should be smaller than any 
transition rate between the system's eigenstates, so that no 
nonresonant transition is induced \cite{messiah,harris}.

We next write Eq. (\ref{s12}) in an integral form and perform the 
integration, 
\[
\sih_{12} = \frac{ g \Eh_1 \sih_{32}^{(1)} -
g \Eh_2^{\dagger} \sih_{13}^{(1)}}{i \ga_c - 2 \De}
\left[ 1 - \alpha e^{-i (2 \De - i \ga_c) t } \right] ,
\]
where $\alpha \simeq [1+ (T_p \De)^2]^{-1}$ is the adiabaticity parameter.
Thus in the adiabatic limit $T_p \gg |\De|^{-1}$, as well as for times
$t \gg \ga_c^{-1}$ (for any $\De$), the term proportional to 
$\alpha$ vanishes. Substituting the above expressions into
\begin{widetext}
\begin{eqnarray*}
\sih_{14} &=&  - \frac{g \Eh_1}{2 \Om_d} -
\frac{g \Eh_2 }{\Om_d} \sih_{12} 
- \frac{i}{\Om_d}
\left[ \left( \ddt - i \de_1 + \frac{\Ga}{2} \right) 
\sih_{13}^{(1)}- \Fh_{13} \right], \\
\sih_{24} &=&  - \frac{g \Eh_2}{2 \Om_d} -
\frac{g \Eh_1 }{\Om_d} \sih_{21} 
- \frac{i}{\Om_d}
\left[ \left( \ddt - i \de_2 + \frac{\Ga}{2} \right) 
\sih_{23}^{(1)}- \Fh_{23} \right], \\
\end{eqnarray*}
after some algebra, we finally arrive at the following set of equations
\begin{subequations}
\label{sslv}
\begin{eqnarray}
\sih_{13} &=& -\frac{i}{\Om_d^{*}} \left[ \ddt -
i(\de_1-\de_d) + \ga_c \right] \sih_{14} + \frac{i}{\Om_d^{*}} \Fh_{14}, 
\label{ss13} \\
\sih_{14} &=& - \frac{g \Eh_1}{2 \Om_d} 
\left[ 1+ \frac{(\de_1+i \Ga/2)(\de_1 - \de_d)}{|\Om_d|^2} 
+ \frac{g^2 \Ih_2 2 \De} 
{|\Om_d|^2 (i \ga_c - 2 \De)} \right] + \frac{i}{\Om_d} \Fh_{13}, 
\label{ss14} \\
\sih_{23} &=& -\frac{i}{\Om_d^{*}} \left[ \ddt  -
i(\de_2-\de_d) + \ga_c \right] \sih_{24} + \frac{i}{\Om_d^{*}} \Fh_{24}, 
\label{ss23} \\
\sih_{24} &=& - \frac{g \Eh_2}{2 \Om_d} 
\left[ 1+ \frac{(\de_2+i \Ga/2)(\de_2 - \de_d)}{|\Om_d|^2}
- \frac{g^2 \Ih_1 2 \De} 
{|\Om_d|^2 (i \ga_c + 2 \De)} \right] + \frac{i}{\Om_d} \Fh_{23},
\label{ss24} 
\end{eqnarray}
\end{subequations}
where $\Ih_j \equiv \Eh_j^{\dagger} \Eh_j$ is the
dimensionless intensity (photon-number) operator for the $j$th field. 

From now on we focus on the case of $\om_p = \om_{31}^{0}= \om_{32}^{0}$.
Substituting Eqs. (\ref{sslv}) into Eqs. (\ref{Es}), the equations of 
motion for quantum fields are obtained as 
\begin{subequations}
\label{Esf}
\begin{eqnarray}
\left[ \ddz + \frac{1}{v_g^{(1)}} \ddt \right] \Eh_1 &=& 
- \ka_1 \Eh_1
-i (\De + \De_d) (s_1 - \eta_1 \Ih_2) \Eh_1 + \Fch_1,  \label{E1s} \\
\left[ \ddz + \frac{1}{v_g^{(2)}} \ddt \right] \Eh_2 &=& 
- \ka_2 \Eh_2
+ i (\De - \De_d) (s_2 - \eta_2 \Ih_1) \Eh_2 + \Fch_2, \label{E2s}
\end{eqnarray}
\end{subequations}
\end{widetext}
where $\De_d=\om_d-\om_{34}^{0} + \De^{\prime}$ is the driving field detuning,
\begin{eqnarray*}
\ka_{1,2} &=& \frac{N g^2}{2 c |\Om_d|^2} 
\left[ \ga_c + \frac{\Ga (\De \pm \De_d)^2}{|\Om_d|^2} \right], \\
s_{1,2} &=& \frac{N g^2}{2 c |\Om_d|^2} 
\left[ 1 + \frac{\De (\De \pm \De_d)}{|\Om_d|^2} \right]  
\end{eqnarray*}
are, respectively, the linear absorption and phase modulation coefficients,
\[
\eta_{1,2} = \frac{N g^4 2 \De}{2 c |\Om_d|^4 (2\De \mp i \ga_c)} 
\]
are the cross-coupling coefficients, $ v_g^{(1,2)} = (1/c + s_{1,2})^{-1}$ 
are the group velocities of the corresponding fields, and $\Fch_{1,2}$ are 
the noise operators having the properties \cite{ScZub}
\begin{eqnarray*}
\expv{\Fch_i(z)} = \expv{\Fch_i(z)\Fch_i(\zp) } &=& 
\expv{\Fch_i^{\dagger}(z)\Fch_i^{\dagger}(\zp) } = 0, \\ 
\expv{\Fch_i(z)\Fch_j^{\dagger}(\zp) } &=& 2 \ka_i \de_{ij} \de(z-\zp) .
\end{eqnarray*}
In deriving Eqs. (\ref{Esf}), we have assumed that the usual EIT conditions 
$|\Om_d|^2 \gg (\De \pm \De_d) k_{p,d} \bar{v}, \ga_c (\Ga + k_{p,d} \bar{v})$,
where $\bar{v}$ is the mean thermal atomic velocity, are satisfied, allowing 
us to neglect the Doppler induced absorption. On the other hand, since the 
terms containing $k_p v$ enter Eqs. (\ref{sslv}) linearly, the net phase-shift 
of the quantum fields, due to the Doppler shifts of the atomic 
resonance frequencies, averages to zero. Note also that if states 
$\ket{1}$, $\ket{2}$ and $\ket{4}$ belong to different hyperfine components 
of a common ground state, the frequencies $\om_p$ and $\om_d$ of the 
optical fields differ from each other by at most a few GHz, 
$\om_p - \om_d \simeq \om_{41}^{0} \ll \om_{p,d}$. Then, as seen 
from Eqs. (\ref{sslv}), the difference $(k_p-k_d) v$ in the Doppler shifts 
of the atomic resonances $\ket{1},\ket{2} \to \ket{3}$ and 
$\ket{4} \to \ket{3}$ is negligible.    

When $\De (\De \pm \De_d) \ll |\Om_d|^2$, the group velocities of 
$\Eh_1$ are $\Eh_2$ are practically the same, 
$v_g^{(1,2)} \simeq v_g$. Expressing the atom-field coupling constant 
$g$ through the linear resonant absorption coefficient 
$a_0 = \wp_{13}^2 \om_p \rho/(\hbar c \eps_0 \Ga)$ for the 
transitions $\ket{1}, \ket{2} \to \ket{3}$ as $N g^2 = a_0 c \Ga/2$ 
and assuming that the density of atoms $\rho = N/(AL)$ is large enough so
that $a_0 c \Ga \gg 4 |\Om_d|^2$, we have 
$v_g \simeq 4 |\Om_d|^2/(a_0 \Ga) \ll c$. Then the solution of 
Eqs. (\ref{Esf}) can be expressed in terms of the retarded time
$\tau = t - z/v_g$ as
\begin{widetext}
\begin{subequations}
\label{Eslv}
\begin{eqnarray}
\Eh_1(z,t) &=& \Eh_1(0,\tau) 
\exp \left[ -\ka_1 z + i \phih_1(z,0,t) \right]
+ \int_0^z d \zp \Fch_1(\zp) 
\exp \left[  -\ka_1 (z-\zp) + i \phih_1(z,\zp,t) \right] , \label{E1slv} \\
\Eh_2(z,t) &=& \Eh_2(0,\tau) 
\exp \left[ -\ka_2 z + i \phih_2(z,0,t) \right]
+ \int_0^z d \zp \Fch_2(\zp) 
\exp \left[  -\ka_2 (z-\zp) + i \phih_2(z,\zp,t) \right] . \label{E2slv}
\end{eqnarray}
\end{subequations}
where the phase operators are given by
\begin{eqnarray*}
\phih_1(z,\zp,t) &=&  - s_1 (\De + \De_d) (z - \zp)
+ \eta_1 (\De + \De_d) 
\int_{\zp}^z d \zpp \Ih_2(\zpp, \tau + \zpp/v_g)  , \; \\
\phih_2(z,\zp,t) &=&  s_2 (\De - \De_d) (z -\zp) 
- \eta_2 (\De - \De_d) 
\int_{\zp}^z d \zpp \Ih_1(\zpp, \tau + \zpp/v_g)  . \; 
\end{eqnarray*}
\end{widetext}
These are the central equations of this paper. The first terms in 
Eqs. (\ref{Eslv}) describe the linear attenuation and the phase shift 
of the corresponding quantum field $\Eh_{1,2}$ upon propagating through 
the medium, while the second terms account for the noise contribution. 
Note that although the expectation values of the field operators decay, 
yet slowly, with the propagation, due to the presence of the noise 
operators, their commutators are preserved \cite{ScZub}. We emphasize 
again that Eqs. (\ref{Eslv}) are obtained within the weak field and 
adiabatic approximations.

In the following section we explore the classical limit of Eqs. (\ref{Eslv}) 
for the purpose of sensitive magnetometry. In Sec. \ref{xpm} we study the 
quantum dynamics of the system and show that our scheme is capable of 
realizing strong nonlinear interaction and entanglement between two tightly 
focused quantum fields at the single-photon level.

\section{Optical magnetometer}
\label{mor}

Let us consider the classical limit of Eqs. (\ref{Eslv}), by replacing the
operators $\Eh_{1,2}$ with the corresponding c-numbers ${\mathcal E}_{1,2}$ 
and dropping the noise terms. The equations for the two circularly polarized 
components of the {\it cw} probe field have the form
\begin{subequations}
\label{Ecwslv}
\begin{eqnarray}
{\mathcal E}_1(z) &=& {\mathcal E}_1(0) e^{-\ka_1 z} e^{i \phi_1(z)}, 
\label{E1cwslv} \\
{\mathcal E}_2(z) &=& {\mathcal E}_2(0) e^{-\ka_2 z} e^{i \phi_2(z)}, 
\label{E2cwslv} 
\end{eqnarray}
\end{subequations}
where the absorption coefficients and phase shifts can be expressed
through the group velocity $v_g$ as
\begin{subequations}
\label{phshfts}
\begin{eqnarray}
\ka_{1,2} &=& \frac{\ga_c}{v_g}+\frac{\Ga (\De \pm \De_d)^2}{v_g |\Om_d|^2}, \\
\phi_1(z) &=& - \frac{\De + \De_d}{v_g} z -
\frac{\De (\De + \De_d)^2}{v_g|\Om_d|^2} z 
\nonumber \\ & &
+ \frac{\De + \De_d}{v_g} \frac{g^2 I_2(0)}{|\Om_d|^2} 
\frac{1-e^{-2 k_2 z}}{2 k_2} , \\
\phi_2(z) &=& \frac{\De - \De_d}{v_g} z +
\frac{\De (\De - \De_d)^2}{v_g|\Om_d|^2} z 
\nonumber \\ & &
- \frac{\De - \De_d}{v_g} \frac{g^2 I_1(0)}{|\Om_d|^2} 
\frac{1-e^{-2 k_1 z}}{2 k_1} .
\end{eqnarray}
\end{subequations}
When the absorption is small, $\ka_{1,2} z \ll 1$, $z \in \{0,L \}$,
which requires that $v_g / \ga_c \gg L$ and 
$\De^2 + \De_d^2 \alt \ga_c |\Om_d|^2/\Ga$, 
the polarization rotation of the probe field 
$\Phi(z) = \frac{1}{2}[\phi_2(z) -\phi_1(z)]$ is given by
\begin{equation}
\Phi(z) = \frac{\De}{v_g}z + 
\frac{\De (\De^2 + \De_d^2)}{v_g |\Om_d|^2}z
+\frac{\De}{v_g}  \frac{g^2 I(0)}{|\Om_d|^2}z , \label{theta}
\end{equation}
where $I(0) = I_1(0) = I_2(0)$ since the probe is linearly polarized at
the entrance to the medium. In Eq. (\ref{theta}), the first term 
is linear in the magnetic field while the second term has a cubic 
dependence on the field strength. Here we focus our attention on 
the measurement of dc magnetic fields employing the dominating linear term.
We wish, however, to point out that the presence of the cubic term
may facilitate the detection of ac fields oscillating slowly compared 
to the bandwidth of the magnetometer, which is limited by the bandwidth
of the EIT window \cite{quantbw}
\begin{equation}
\de \om \leq \frac{|\Om_d|^2}{\Ga} \frac{k_p}{\sqrt{3 \pi \rho L}} . 
\label{eit_bw}
\end{equation}
Then the spectrum of $\Phi$, along with the fundamental frequency of the 
magnetic field, will also contain its third harmonic which, for very small 
frequencies, may be easier to detect \cite{kominis}. This issue is beyond 
the scope of this paper and will be addressed elsewhere. Finally, the last 
term of Eq. (\ref{theta}), being proportional to the product of the magnetic 
field strength and probe intensity, is a consequence of Kerr-type 
nonlinear interaction between ${\mathcal E}_1$ and ${\mathcal E}_2$, 
which is the subject of the following section. 

We consider a magnetometer setup in the ``balanced polarimeter'' arrangement 
\cite{budker_rev}, in which, at the exit from the medium $z=L$, a polarizing 
beam splitter oriented at $\pi/4$ to the input polarizer [$\Phi(0)=0$] is 
used as an analyzer. Then the detector signal $S$ is represented by
the difference of photocounts in the two channels of the analyzer
\begin{equation}
S = 2 n_{\rm in} e^{-2 \ka L} \sin[\Phi(L)] \cos[\Phi(L)] , \label{signal}
\end{equation}
where $n_{\rm in} = P_{\rm in} t_m/(\hbar \om_p) = 2 I(0) c t_m/L$, 
with $P_{\rm in}$ being the input power of the probe, is the number 
of photons passing through the medium during the measurement time $t_m$.
For simplicity, we neglect the difference between the absorption 
coefficients for the circularly left- and right-polarized components, 
$\ka_1 \simeq \ka_2 = \ka$, which amounts to neglecting the ellipticity 
of the output field ($\veps \simeq 1$) since 
$\sqrt{1 - \veps^2} = 2\Ga \De \De_d L /(v_g |\Om_d|^2) \ll 1$.

The most important characteristic of a magnetometer is its {\it sensitivity}
to weak magnetic fields, which is limited by the measurement noise. The 
smallest detectable magnetic field $B_{\rm min}$ can be defined as being 
the field for which the signal is equal to the noise. In our system, 
the total noise ${\cal N} = {\cal N}_{\rm at} + {\cal N}_{\rm shot}$ 
has two contributions, atomic noise ${\cal N}_{\rm at}$ and photon counting 
shot-noise ${\cal N}_{\rm shot}$. The atomic contribution is due to the 
spontaneous photons reaching the detector during the measurement time,
\[
{\cal N}_{\rm at} = \Ga \expv{\sih_{33}} N t_m \frac{A}{4 \pi L^2},  
\]  
where the detector area is assumed to be equal to $A$. For vanishing 
magnetic field $\De < \ga_c$, we have  
$\expv{\sih_{33}} N \simeq a_0 c \Ga \ga_c^2I(0) /(4 |\Om_d|^4)$
and the atomic noise is given by 
\[
{\cal N}_{\rm at} = \frac{a_0 \Ga^2 \ga_c^2 A}{32 \pi |\Om_d|^4 L} n_{\rm in}.
\]
For physically realistic parameters (see below), the atomic noise term 
is small compared to the photon counting shot-noise \cite{flscl}
\[
{\cal N}_{\rm shot} = \sqrt{\frac{1+e^{-2 \ka L}}{2}  n_{\rm in}} \leq
 \sqrt{n_{\rm in}}.
\]
In the limit of weak magnetic field, retaining only the linear in magnetic 
field term, from $S \simeq 2 n_{\rm in} \Phi \geq  {\cal N}_{\rm shot}$ 
we obtain
\begin{equation}
B_{\rm min} \geq \frac{2 \hbar |\Om_d|^2}
{g_F \mu_B a_0 L \Ga \sqrt{n_{\rm in}}}.
\end{equation}
For realistic experimental parameters, $\om_p = 3 \times 10^{15}$ rad/s,
$\Ga = 10^7$ s$^{-1}$, $\rho = 10^{13}$ cm$^{-3}$ 
($a_0 \simeq 10^{4}$ cm$^{-1}$), $|g_F| = 1/2$, $\Om_d \simeq \Ga$, 
$L = 10$ cm, $P_{\rm in} = 1$ mW, $t_m = 1$ s, the minimum detectable
magnetic field $B_{\rm min} \lesssim 10^{-12}$ G, which is of the
same order as that of \cite{butker,scully-mor,flscl}. 
Thus, concerning the magnetometer sensitivity, our scheme is 
essentially equivalent to the one proposed in \cite{flscl}, where
an interferometric measurement of the magnetic field induced phase 
shift of a probe field, subject to EIT with $\La$ atoms, was studied.
Experimentally, however, measuring the polarization rotation of the 
probe, as suggested here, may be more practical than measuring 
its phase shift in the setup of \cite{flscl}, which employs a 
Mach-Zehnder interferometer.

\section{Cross-phase modulation}
\label{xpm}

In order to rigorously describe the nonlinear interaction between the 
weak {\it pulsed} fields, we now turn to the fully quantum treatment 
of the system. When absorption is small enough to be neglected,
from Eqs. (\ref{Eslv}) we have 
\begin{subequations}
\label{qEslv}
\begin{eqnarray}
\Eh_1(z,t) &=& \Eh_1(0,\tau) 
\exp [ i \eta (\De + \De_d)  \Eh_2^{\dagger}(0,\tau) 
\Eh_2(0,\tau) z ], 
\nonumber \\ \label{qE1slv} \\
\Eh_2(z,t) &=& \Eh_1(0,\tau) 
\exp [ - i \eta (\De - \De_d)  \Eh_1^{\dagger}(0,\tau) 
\Eh_1(0,\tau) z ], 
\nonumber \\ \label{qE2slv} 
\end{eqnarray}
\end{subequations}
where the cross-phase modulation coefficient is given by 
$\eta = g^2/(v_g |\Om_d|^2)$ (assuming $\ga_c \ll \De$), 
while the linear phase-modulation is 
incorporated into the field operators via the unitary transformations
$\Eh_1(z,t) \to \Eh_1(z,t) e^{- i s_1 (\De + \De_d)z}$ and 
$\Eh_2(z,t) \to \Eh_2(z,t) e^{i s_2 (\De - \De_d)z}$.
These traveling-wave electric fields can be expressed through 
single mode operators as 
$\Eh_j(z,t) = \sum_{q} a_{j}^{q}(t) e^{i q z}$ ($j=1,2$), 
where $a_j^q$ is the annihilation operator for the field mode 
with the wavevector $k_p + q$. The single-mode operators $a_j^q$ and 
$a^{q \dagger}_j$ possess the standard bosonic commutation relations 
$[a_{i}^{q},a_{j}^{q^{\prime}\dagger}]=\de_{ij} \de_{qq^{\prime}}$.
The continuum of modes scanned by $q \in \{-\de q/2, \de q/2\}$ is bounded
by the EIT window via $\de q \leq \de \om /c$ \cite{lukimam}. The finite 
quantization bandwidth $\de q$ for the field operators leads to the 
equal-time commutation relations
\[
[\Eh_{i}(z),\Eh_{j}^{\dagger}(\zp)]= 
\de_{ij} \frac{L \de q}{2 \pi} {\rm sinc}\left[ \de q (z-\zp)/2 \right] ,
\]
where ${\rm sinc}(x) = \sin(x)/x$.  

Before proceeding, we note that Eqs. (\ref{qEslv}) are similar to the 
corresponding equations of Ref. \cite{lukimam}, where the cross-phase 
modulation between two quantum fields was mediated by atoms with N 
configuration of levels \cite{imam}, while the group velocity mismatch 
between the fields was compensated by using a second kind of 
$\La$-atoms controlled by an additional driving field. 
In contrast, our scheme relies solely on an intra-atomic process 
employing only one driving field that causes simultaneous EIT for 
both fields and their cross-coupling. It is therefore deprived of 
complications associated with using mixtures of two isotopic species 
of atoms \cite{lukimam} or invoking cavity QED techniques \cite{phphcav}.

The most classical of all the quantum states is the coherent state.
To compare the classical and quantum pictures, we therefore consider 
first the evolution of input wavepacket
$\ket{\psi_{\rm in}} = \ket{\alpha_1} \otimes \ket{\alpha_2}$ composed of the 
multimode coherent states $\ket{\alpha_j} \equiv \Pi_{q} \ket {\alpha_j^q}$
($j=1,2$). The states $\ket{\alpha_j}$ are the eigenstates of the input 
operators $\Eh_j(0,t)$ at $z=0$ with the eigenvalues
$\alpha_j(t) = \sum_{q} \alpha_j^{q} e^{-i q c t}$:
$\Eh_j(0,t)\ket{\alpha_j} = \alpha_j(t) \ket{\alpha_j}$.
Upon propagating through the medium, each pulse experiences a nonlinear 
cross-phase modulation. The expectation values for the fields are then 
obtained as
\begin{subequations}
\label{expval}
\begin{eqnarray}
\expv{\Eh_1 (z,t)} &=& \alpha_1(\tau)  \exp \left\{
\left[ e^{i \theta_1(z)}-1 \right] 
\frac{2 \pi |\alpha_2(\tau)|^2}{L \de q} \right\} , \;\;\;\;\;  \\
\expv{\Eh_2 (z,t)} &=& \alpha_2(\tau)  \exp \left\{
\left[ e^{i \theta_2(z)}-1 \right] 
\frac{2 \pi |\alpha_1(\tau)|^2}{L \de q} \right\} ,
\end{eqnarray}
\end{subequations}
where $\theta_{1,2}(z) = \eta (\De_d \pm \De) L \de q z/(2 \pi)$. 
These equations are similar to those obtained for single-mode 
\cite{sami} and multimode copropagating fields \cite{lukimam}. 
They indicate that when the cross-phase modulation is large, 
upon propagating through the medium, the phases 
\[
2 \pi \sin[\theta_{1,2}(z)]\frac{|\alpha_{2,1}(\tau)|^2}{L \de q}
\]
and amplitudes
\[
\alpha_{1,2}(\tau) \exp \left\{- 4 \pi \sin^2 [\theta_{1,2}(z)/2 ]
\frac{|\alpha_{2,1}(\tau)|^2}{L \de q} \right\}
\]
of the quantum fields exhibit periodic collapses and revivals as 
$\theta_{1,2}(z)$ change from 0 to $2 \pi$. In particular, when the 
phase-shift is maximal, $\theta_{1,2} = \pi/2$, the amplitude of the
corresponding field is reduced by a factor of 
$r_{1,2} = \exp [- 2 \pi |\alpha_{2,1}|^2/(L \de q)]$. On the 
other hand, the maximal dephasing of the multimode coherent field,
$r_{1,2} = \exp [- 4 \pi |\alpha_{2,1}|^2/(L \de q)]$, 
is attained for $\theta_{1,2}(z) = (2 n +1)\pi$ ($n=0,1,2,\ldots$),
where the phase shift is zero. 

We have thus seen that the behavior of weak quantum fields is remarkably 
different from that of classical fields, as in the quantum regime the 
nonlinear phase shift is bounded between 
$\pm 2 \pi |\alpha_{2,1}|^2/(L \de q)$. Only in the limit of weak 
cross-phase modulation $\theta_{1,2} \ll 1$, the quantum Eqs. (\ref{expval}) 
reproduce the classical result
\[
\expv{\Eh_{1,2} (z,t)} = \alpha_{1,2}(\tau)  
\exp [ i \eta (\De_d \pm \De) |\alpha_{2,1}(\tau)|^2 z] ,
\]
whereby the cross-phase shift grows linearly with the propagation
distance and can attain large values when the field amplitudes are
sufficiently high.

Let us now consider the input state 
$\ket{\psi_{\rm in}} = \ket{1_1} \otimes \ket{1_2}$, consisting of two 
single photon wavepackets $\ket{1_{j}} = \sum_{q} \xi_{j}^q 
a_{j}^{q \dagger} \ket{0}$ ($j = 1,2$). The Fourier amplitudes 
$\xi_j^q$, normalized as $\sum_{q} |\xi_{j}^q|^2 =1$, define the 
spatial envelopes $f_{j}(z)$ of the two pulses that initially 
(at $t=0$) are localized around $z=0$,
\[
\bra{0} \Eh_{j}(z,0) \ket{1_j} = \sum_{q} \xi_{j}^q e^{iqz} = f_{j}(z).
\]
In free space, $\Eh_{j}(z,t) =\Eh_{j}(0,\tau)$ with $\tau = t -z/c$,
and we have $\bra{0} \Eh_{j}(z,t) \ket{1_j} = f_{j}(z-ct)$.  
The state of the system at any time can be represented as
\begin{equation}
\ket{\psi(t)} = \sum_{q,q^{\prime}} \xi_{12}^{qq^{\prime}}(t)
\ket{1_1^q} \ket{1_2^{q^{\prime}}}, \label{st}
\end{equation}
from where it is apparent that 
$\xi_{12}^{qq^{\prime}}(0) = \xi_{1}^q \xi_{2}^{q^{\prime}}$.

Since for the photon-number states the expectation values of the 
field operators vanish, all the information about the state of the system
is contained in the intensities of the corresponding fields 
\begin{equation}
\expv{\Ih_j (z,t)} = \bra{\psi_{\rm in}} \Eh_{j}^{\dagger} (z,t) 
\Eh_j (z,t)\ket{\psi_{\rm in}} , \label{evI}
\end{equation}
and their ``two-photon wavefunction'' \cite{ScZub,lukimam}
\begin{equation}
\Psi_{ij}(z,t;\zp,t^{\prime}) = 
\bra{0} \Eh_j(\zp,t^{\prime}) \Eh_i(z,t) \ket{\psi_{\rm in}}.
\label{tphwf}
\end{equation}
The physical meaning of $\Psi_{ij}$ is a two-photon detection amplitude,
through which one can express the second-order correlation function
$G^{(2)}_{ij} = \Psi_{ij}^{*} \Psi_{ij}$ \cite{ScZub}. The knowledge 
of the two-photon wavefunction allows one to calculate the amplitudes 
$\xi_{12}^{qq^{\prime}}$ of state vector (\ref{st}) via the two 
dimensional Fourier transform of $\Psi_{ij}$ at $t = t^{\prime}$:
\begin{equation}
\xi_{ij}^{qq^{\prime}}(t) = \frac{1}{L^2} \int \!\!\! \int dz d \zp 
\Psi_{ij}(z,\zp,t) e^{-iqz}e^{-iq^{\prime} \zp } \label{ftrns}.  
\end{equation}

We first calculate the expectation values of the intensities 
$\expv{\Ih_j (z,t)}$ by substituting the operator solution (\ref{qEslv}) 
into (\ref{evI}), 
\begin{equation}
\expv{\Ih_j (z,t)} = |f_j(-c \tau)|^2 = |f_j(z c /v_g - ct)|^2 , 
\label{res-evI}
\end{equation}
where $\tau = t - z/v_g$ for $0 \leq z < L$. This equation indicates that
upon entering the medium, as the group velocities of the pulses are slowed 
down to $v_g \ll c$, their spatial envelopes are compressed by a factor of 
$c/v_g$ \cite{fllk}. Outside the medium, at $z \geq  L$ and accordingly 
$\tau = t - L/v_g -(z-L)/c$, we have 
$\expv{\Ih_j (z,t)} = |f_j(z + L(c/v_g -1) -ct)|^2$, which shows that
the propagation velocity and the pulse envelopes are restored to their
free-space values. 

Consider next the two photon wavefunction $\Psi_{ij}$. After the interaction,
at $z,\zp \geq L$, we have the general expression
\begin{widetext}
\begin{eqnarray}
\Psi_{ij}(z,t; \zp ,t^{\prime}) &=& f_i(-c \tau) f_j(-c \tau^{\prime})
\left\{ 1 + \frac{f_j(-c \tau)}{f_j(-c \tau^{\prime})} 
{\rm sinc}\left[ \frac{\de \om}{2} (\tau- \tau^{\prime}) \right]
\left(e^{i \theta_i(L)} -1 \right) \right\} , \label{res-tphwf}
\end{eqnarray}
where, as before, $\tau = t-L/v_g-(z-L)/c$ and similarly for $\tau^{\prime}$.
For quantum information applications, it makes sense to consider the
relatively simple case of small magnetic field, such that 
$\De,\De^{\prime} \ll \De_d$, where the driving field detuning 
$\De_d=\om_d-\om_{34}^{0}$ satisfies $|\De_d| < \de \om /2$. 
We thus have $\theta_{1,2} \simeq \theta = \eta \De_d L^2 \de q /(2 \pi)$.
Then the equal-time ($t = t^{\prime}$) two-photon wavefunction reads 
\begin{eqnarray}
\Psi_{ij}(z,\zp ,t) &=& f_i[z +L(c/v_g -1) -c t] \:
f_j[\zp + L(c/v_g -1) -c t]
\nonumber  \\ & & \times 
\left\{ 1 + 
\frac{f_j[z+ L(c/v_g -1) -c t]}{f_j[ \zp + L(c/v_g -1) -c t]} \:
{\rm sinc}\left[ \frac{\de q}{2} (\zp - z) \right]
\left(e^{i \theta} -1 \right) \right\} . \label{et-tphwf}
\end{eqnarray}
\end{widetext}
For large enough spatial separation between the two photons, 
such that $|\zp - z| > \de q^{-1}$ and therefore
${\rm sinc} [ \de q (\zp - z)/2] \simeq 0$, Eq. (\ref{et-tphwf})
yields
\[
\Psi_{ij}(z,\zp ,t) \simeq  
f_i[z +L(c/v_g -1) -c t] \: f_j[\zp + L(c/v_g -1) -c t] ,
\]
which indicates that no nonlinear interaction takes place between 
the photons, which emerge from the medium unchanged. This is due 
to the {\it local} character of the interaction described by the 
${\rm sinc}$ function. 

Consider now the opposite limit of $|\zp - z| \ll \de q^{-1}$ and 
therefore ${\rm sinc} [ \de q (\zp - z)/2] \simeq 1$. Then for two 
narrow-band (Fourier limited) pulses with the duration 
$T_p \gg |\zp - z|/c$, one has $f_j(z)/f_j(\zp ) \simeq 1$, 
and Eq. (\ref{et-tphwf}) results in 
\begin{eqnarray*}
\Psi_{ij}(z,\zp ,t) &\simeq & e^{i \theta}
f_i[z +L(c/v_g -1) -c t] 
\nonumber  \\ & & \;\;\;\; \times 
f_j[\zp + L(c/v_g -1) -c t] .
\end{eqnarray*}
Thus, after the interaction, a pair of single photons acquires conditional
phase shift $\theta$, which can exceed $\pi$ when 
\[
\left( \frac{\de q L}{2 \pi} \right)^2 > \frac{v_g |\Om_d|^2}{c g^2}.
\]
To see this more clearly, we use Eq. (\ref{ftrns}) to calculate the
amplitudes of the state vector $\ket{\psi(t)}$:
\begin{equation}
\xi_{ij}^{qq^{\prime}}(t) = e^{i \theta}  \xi_{ij}^{qq^{\prime}}(0)  
\exp \{i (q+q^{\prime}) [L(c/v_g -1) -ct ] \}. \label{res-ftrns}
\end{equation}
At the exit from the medium, at time $t \simeq L/v_g$, the second exponent 
in Eq. (\ref{res-ftrns}) can be neglected for all $q,q^{\prime}$ and the 
state of the system is given by 
\begin{equation}
\ket{\psi(L/v_g)} = e^{i \theta}\ket{\psi_{\rm in}}.
\end{equation} 
When $\theta = \pi$, this transformation corresponds to the truth table 
of the {\it controlled-phase} (\textsc{cphase}) logic gate between
the two photons representing qubits. Together with the linear
single-photon phase shifts (realizing single-qubit rotations), 
the \textsc{cphase} gate is said to be {\it universal} in the sense
that it can realize arbitrary unitary transformation \cite{QCQI}. 

\section{Conclusions}
\label{sum}

In this paper, we have studied a propagation of weak probe field  
through an optically dense medium of coherently driven four-level
atoms in a tripod configuration. We have presented a detailed semiclassical 
as well as quantum analysis of the system. One of the conclusions that 
emerged from this study is that optically dense vapors of tripod atoms 
can support ultrasensitive magneto-optical polarization rotation of the 
probe field and therefore have significant potential for improving the 
sensitivity of Faraday magnetometers to small magnetic fields. Another 
finding is that this system is capable of realizing a novel regime of 
symmetric, extremely efficient nonlinear interaction of two multimode 
single-photon pulses, whereby the combined state of the system 
acquires a large conditional phase shift that can easily exceed $\pi$. 
Thus our scheme may pave the way to photon-based quantum information 
applications, such as {\it deterministic} all-optical quantum computation, 
dense coding and teleportation \cite{QCQI}.


\begin{thebibliography}{99}

\bibitem{eit_rev} S.E. Harris, Phys. Today {\bf 50}(7), 36 (1997);
E. Arimondo, in {\it Progress in Optics}, edited by E. Wolf,  
(Elsevier Science, Amsterdam, 1996), vol. 35, p. 257.

\bibitem{ScZub}
M.O.~Scully and M.S.~Zubairy, {\em Quantum Optics}
(Cambridge University Press, Cambridge, UK, 1997).

\bibitem{vred} L.V.~Hau, S.E.~Harris, Z.~Dutton, and C.H.~Behroozi,
Nature (London) {\bf 397}, 594 (1999);
M.M.~Kash, V.A.~Sautenkov, A.S.~Zibrov, L.~Hollberg, G.R.~Welch, 
M.D.~Lukin, Yu.~Rostovtsev, E.S.~Fry, and M.O.~Scully,
Phys. Rev. Lett. {\bf 82}, 5229 (1999); 
D.~Budker, D.F.~Kimball, S.M.~Rochester, and V.V.~Yashchuk, 
Phys. Rev. Lett. {\bf 83}, 1767 (1999).

\bibitem{acopt} 
A.B.~Matsko, Yu.V.~Rostovtsev, H.Z.~Cummins, and M.O.~Scully,
Phys. Rev. Lett. {\bf 84}, 5752 (2000).

\bibitem{fllk}
M.~Fleischhauer and M.D.~Lukin, Phys. Rev. Lett. {\bf 84}, 5094 (2000);
Phys. Rev. A {\bf 65}, 022314 (2002).

\bibitem{v0exp} 
D.F.~Phillips, A.~Fleischhauer, A.~Mair, R.L.~Walsworth, and M.D.~Lukin,
Phys. Rev. Lett. {\bf 86}, 783 (2001);
C.~Liu, Z.~Dutton, C.H.~Behroozi, and L.V.~Hau,
Nature {\bf 409}, 490 (2001).

\bibitem{hemmer}  A.V.~Turukhin, V.S.~Sudarshanam, M.S.~Shahriar, 
J.A.~Musser, B.S.~Ham, and P.R.~Hemmer, 
Phys. Rev. Lett. {\bf 88}, 023602 (2002).

\bibitem{lukin-pbg}
A.~Andre and M.D.~Lukin, Phys. Rev. Lett. {\bf 89}, 143602 (2002);
M.~Bajcsy, A.S.~Zibrov, M.D.~Lukin,
Nature {\bf 426}, 638 (2003).

\bibitem{imam}
H.~Schmidt and A.~Imamo\u{g}lu, Opt. Lett. {\bf 21}, 1936 (1996).

\bibitem{harris}
S.E.~Harris and Y.~Yamamoto, Phys. Rev. Lett. {\bf 81}, 3611 (1998);
S.~Harris and L.~Hau, Phys. Rev. Lett. {\bf 82}, 4611 (1999).

\bibitem{bergmann}
F.~Vewinger, M.~Heinz, R.G.~Fernandez, N.V.~Vitanov, and K.~Bergmann,
Phys. Rev. Lett. {\bf 91}, 213001 (2003).

\bibitem{bergmann-rev} 
K.~Bergmann, H.~Theuer, and B.W.~Shore,
Rev. Mod. Phys. {\bf 70}, 1003 (1998).

\bibitem{pasp-knight} 
E.~Paspalakis, N.J.~Kylstra, and P.~Knight,
Phys. Rev. A {\bf 65}, 053808 (2002);
E.~Paspalakis and P.~Knight,
J Mod. Opt. {\bf 49}, 87 (2002).

\bibitem{kis-pasp} E.~Paspalakis and Z.~Kis,
Opt. Lett. {\bf 27}, 1836 (2002);
Z.~Kis and E.~Paspalakis,
Phys. Rev. A {\bf 68}, 043817 (2003).

\bibitem{yumal}  Yu.P.~Malakyan, quant-ph/0112058.

\bibitem{budker_rev}  D.~Budker, W.~Gawlik, D.F.~Kimball, S.M.~Rochester, 
V.V.~Yashchuk, and A.~Weis, Rev. Mod. Phys. {\bf 74}(4), 1153 (2002).

\bibitem{butker}  D.~Budker, V.~Yashchuk, and M.~Zolotorev, 
Phys. Rev. Lett. {\bf 81}, 5788 (1998);
D.~Budker, D F.~Kimball, S.M.~Rochester, V.V.~Yashchuk, 
and M.~Zolotorev, Phys. Rev. A {\bf 62}, 043403 (2000).

\bibitem{scully-mor}  V.A.~Sautenkov, M.D.~Lukin, C.J.~Bednar, I.~Novikova, 
E.~Mikhailov, M.~Fleischhauer, V.L.~Velichansky, G.R.~Welch, and M.O.~Scully,
Phys. Rev. A {\bf 62}, 023810 (2000);
I.~Novikova, A.B.~Matsko, V.A.~Sautenkov, V.L.~Velichansky, G.R.~Welch, 
and M.O.~Scully, Opt. Lett. {\bf 25}, 1651 (2000);
I.~Novikova, A.B.~Matsko, and G.R.~Welch, 
Opt. Lett. {\bf 26}, 1016 (2001).

\bibitem{flscl}  
M.O.~Scully and M.~Fleischhauer, 
Phys. Rev. Lett. {\bf 69}, 1360 (1992);
M.~Fleischhauer and M.O.~Scully, 
Phys. Rev. A {\bf 49}, 1973 (1994).

\bibitem{opt-pump} 
J.~DuPont-Roc, S.~Haroche, and C.~Cohen-Tannoudji,
Phys. Lett. A {\bf 28}, 638 (1969);
I.~Kominis, T.~Kornack, J.~Allred, and M.~Romalis,
Nature (London) {\bf 422}, 596 (2003).

\bibitem{QCQI}
M.A.~Nielsen and I.L.~Chuang, {\em Quantum Computation and Quantum
  Information} (Cambridge University Press, Cambridge, UK, 2000); 
A.~Steane, Rep. Prog. Phys. {\bf 61}, 117 (1998);
C.H.~Bennett and D.P.~DiVincenzo, Nature {\bf 404}, 247 (2000).

\bibitem{solst}
D.~Loss and D.P.~DiVincenzo, Phys. Rev. A {\bf 57}, 120 (1998);
B.E.~Kane, Nature {\bf 393}, 133 (1998).

\bibitem{iontr}
J.I.~Cirac and P.~Zoller, Phys. Rev. Lett. {\bf 74}, 4091 (1995);
C.A.~Sackett, D.~Kielpinski, B.E.~King, C.~Langer, V.~Meyer, 
C.J.~Myatt, M.~Rowe, Q.A.~Turchette, W.M.~Itano, D.J.~Wineland, 
and C.~Monroe, Nature {\bf 404}, 256 (2000).

\bibitem{BCJD}
G.K.~Brennen, C.M.~Caves, P.S.~Jessen, and I.H.~Deutsch, 
Phys. Rev. Lett. {\bf 82}, 1060 (1999).

\bibitem{linopt}
E.~Knill, R.~Laflamme, and G.J.~Milburn, Nature {\bf 409}, 46 (2001);
L.-M.~Duan, M.D.~Lukin, J.I.~Cirac, and P.~Zoller, 
Nature {\bf 414}, 413 (2001).

\bibitem{phphcav}
Q.A.~Turchette, C.J.~Hood, W.~Lange, H.~Mabuchi, and H.J.~Kimble, 
Phys. Rev. Lett. {\bf 75}, 4710 (1995);
A.~Imamoglu, H.~Schmidt, G.~Woods, and M.~Deutsch, 
Phys. Rev. Lett. {\bf 79}, 1467 (1997); 
A.~Rauschenbeutel, G.~Nogues, S.~Osnaghi, P.~Bertet, M.~Brune, 
J.M.~Raimond, and S.~Haroche, Phys. Rev. Lett. {\bf 83}, 5166 (1999).

\bibitem{Boyd}
R.W.~Boyd, {\em Nonlinear Optics} (Academic Press, San Diego, CA, 1992).

\bibitem{lukimam}
M.D.~Lukin and A.~Imamo\u{g}lu, Phys. Rev. Lett. {\bf 84}, 1419 (2000).

\bibitem{rebic}  S.~Rebic, D.~Vitali, C.~Ottaviani, P.~Tombesi, 
M.~Artoni, F.~Cataliotti, and R.~Corbalan, quant-ph/0310148.

\bibitem{messiah} A. Messiah, {\em Quantum Mechanics}
(Elsevier Science, New York, 1981).

\bibitem{quantbw}
M.D.~Lukin, M.~Fleischhauer, A.S.~Zibrov, H.G.~Robinson, V.L.~Velichansky, 
L.~Hollberg, and M.O.~Scully, Phys. Rev. Lett. {\bf 79}, 2959 (1997).

\bibitem{kominis} I.~Kominis (private communications).

\bibitem{sami}
B.C.~Sanders and G.J.~Milburn, Phys. Rev. A {\bf 45}, 1919 (1992).

\end{thebibliography}
\end{document}